\documentclass[preprint2]{aastex}

\newcommand {\hii}{H\,{\sc ii}} 
\newcommand {\kms}{\relax \ifmmode {\,\rm km\,s}^{-1}\else \,km\,s$^{-1}$\fi}

\newcommand {\hI}{\rm HI}

\newcommand {\hd}{HD\,5980}

\shorttitle{NGC\,346}
\shortauthors{Naz\'e et al.}

\begin{document}

\title{An X-ray investigation of the NGC\,346 field in the SMC (2) : the field population}

\author{Y. Naz\'e\altaffilmark{1,2}, J.M. Hartwell\altaffilmark{3},
I.R. Stevens\altaffilmark{3}, J. Manfroid\altaffilmark{1,4,5},
S. Marchenko\altaffilmark{6}, M. F. Corcoran\altaffilmark{7},
A.F.J. Moffat\altaffilmark{8}, G. Skalkowski\altaffilmark{8}} 

\altaffiltext{1}{Institut d'Astrophysique et de G\'eophysique,
Universit\'e de Li\`ege, All\'ee du 6 Ao\^ut 17, Bat. B5c, B 4000 -
Li\`ege (Belgium); naze@astro.ulg.ac.be, manfroid@astro.ulg.ac.be} 
\altaffiltext{2}{Research Fellow F.N.R.S.}
\altaffiltext{3}{School of Physics \& Astronomy, University of
Birmingham, Edgbaston, Birmingham B15 2TT (UK); jmh@star.sr.bham.ac.uk,
irs@star.sr.bham.ac.uk} 
\altaffiltext{4}{Research Director F.N.R.S.}
\altaffiltext{5}{Visiting Astronomer, European Southern Observatory}
\altaffiltext{6}{Department of Physics and Astronomy,
Thompson Complex Central Wing, Western Kentucky University,
Bowling Green, KY 42101-3576 (USA), sergey@astro.wku.edu }
\altaffiltext{7}{Universities Space Research Association, High Energy
Astrophysics Science Archive Research Center, Goddard Space Flight
Center, Greenbelt, MD 20771; corcoran@barnegat.gsfc.nasa.gov} 
\altaffiltext{8}{D\'epartement de physique, Universit\'e de Montreal,
C.P. 6128, Succ. Centre-Ville, Montreal, QC, H3C 3J7 (Canada);
 moffat@astro.umontreal.ca, gwen@astro.umontreal.ca} 

\begin{abstract}

We present results from a {\it Chandra} observation of the NGC\,346 
cluster, which is the ionizing source of N66, the most luminous \hii\ region 
and the largest star formation region in the SMC. In the first part 
of this investigation, we have analysed the X-ray properties of the
cluster itself and the remarkable star \hd. But the field contains 
additional objects of interest. In total, 75 X-ray point sources were 
detected in the {\it Chandra} observation: this is five times 
the number of sources detected by previous X-ray surveys. We
investigate here their characteristics in detail. Due to high foreground
absorption, the sources possess rather high hardness ratios.  
Their cumulative luminosity function appears generally steeper 
than that for the rest of 
the SMC at higher luminosities. Their absorption columns suggest that 
most of the sources belong to NGC\,346. Using DSS data and new $UBVRI$ 
imaging with the ESO 2.2m telescope, we also discovered possible 
counterparts for 32 of these X-ray sources and estimated a B spectral type 
for a large number of these counterparts. This tends to suggest that 
most of the X-ray sources in the field are in fact X-ray binaries. 
Finally, some objects show X-ray and/or optical variability, with a
need for further monitoring.
\end{abstract}


\section{Introduction}

The launch of the {\it Chandra} satellite provides an opportunity 
to explore the X-ray sky with a far greater sensitivity and spatial 
resolution than ever before. In a given field of interest, these
characteristics enable the discovery of numerous X-ray sources in 
addition to the main target(s). Although often regarded as secondary,
these sources provide important information which can improve 
our knowledge of the X-ray emission mechanisms in the Universe.\\

These serendipitous discoveries also enable us to study the source 
distribution in galaxies, especially Magellanic dwarf galaxies in our 
case. Most of the dwarf galaxies studied are at a distance of a few Mpc 
(e.g. Martin, Kobulnicky, \& Heckman 2002), and their X-ray observations 
only sample the 
population with L$_X$ $>$ 10$^{36}$ erg s$^{-1}$. On the other hand, 
the relative closeness of the Magellanic Clouds enable us to probe 
the X-ray sources with luminosities as low as 10$^{32}$ erg s$^{-1}$. 
Putting together the informations on faint and bright X-ray sources 
will ultimately lead to a better understanding of this type of galaxies.\\

We have obtained a deep {\it Chandra} observation of the giant \hii\
region N66 \citep{he56}, the largest star formation region in the Small
Magellanic Cloud (SMC).  
The large number of massive stars and the presence of the remarkable 
star \hd\ make the NGC\,346 field one of the best opportunities to 
conduct an investigation of the X-ray domain. \\

In the first part of this analysis (Naz\'e et al. 2002, paper I), 
we have presented the characteristics of the cluster,
the star \hd\ and its close neighborhood. The cluster itself is 
relatively faint and most of its emission seems correlated with the 
location of the brightest stars in the core. However, the level of 
X-ray emission could not be explained solely by the emission from 
individual stars. The {\it Chandra} observation also provides the 
first X-ray detection of \hd. In X-rays, the star, that underwent a 
LBV-type eruption in 1994, appears very bright, comparable only 
to the brightest WR stars in the Galaxy. This high luminosity could 
be explained either by colliding winds in the binary system or by 
post-eruption effects. Finally, a bright, extended X-ray emission 
seems to surround this star. It is probably due to a SNR which may 
or may not be related to \hd\ itself (see paper I). \\

In this second paper, we will focus on the other X-ray sources present 
in the field. First, we will describe in \S~2 the observations 
used in this study. The detected sources, their hardness ratios (HRs), 
and their spectral 
characteristics will then be discussed in \S~3, 4, and 5, respectively. 
Next, we will describe the overall properties of the point sources' 
population in \S~6, present their possible counterparts in \S~7 and 
investigate their variability in \S~8. Finally, we will give a summary in 
\S~9.\\

\section{The Observations and Data Analysis}

\subsection{X-ray Observations}

NGC\,346 was observed with {\it Chandra} for the 
XMEGA\footnote{http://lheawww.gsfc.nasa.gov/users/corcoran/xmega/xmega.html} 
consortium on 2001 May 15--16 for 100 ks (98.671 ks effective, ObsID = 1881,
JD$\sim$2\,452\,045.2).  The data were recorded by the ACIS-I0, I1, I2,
I3, S2 and S3 CCD chips maintained at a temperature of
$-120^\circ$C. The data were taken with a frame time of 3.241s in the
faint mode. Our faintest
sources have fluxes of about $6\times 10^{32}$~erg~s$^{-1}$ (see
\S 6), assuming a distance of 59~kpc for the SMC \citep{ma86}. More details
about the processing of the data can be found in paper I.\\

For long exposures, removing the afterglow events can adversely affect
the science analysis (underestimation of the fluxes, alteration of the
spectra and so on, see paper I). We thus computed a new level 2 events 
file and we 
will use this new file throughout this paper, except for the source 
detection, where it is better to use the pipeline level 2 product to 
avoid mistaking afterglow events for real sources.\\

Further analysis was performed using the CIAO v2.1.2 software provided
by the CXC and also with the FTOOLS tasks. The spectra were analysed and
fitted within XSPEC v11.0.1 \citep{ar96}. Fig. \ref{totalfield} shows 
an image of the four ACIS-I chips : numerous point sources and some 
diffuse emission near \hd\ can be seen. Fig. \ref{dss} presents
the superposition of the {\it Chandra} data on a DSS image.
The ACIS-I field-of-view covers
290~pc$\times$~290~pc, while the size of NGC\,346 is 180~pc$\times$220~pc
\citep{ye91}. Note that due to their high noise and poor resolution, the 
ACIS-S data won't be analysed in details here.\\

\subsection{WFI Images}

Optical imaging data were acquired as part of a project 
aiming at the photometric calibration of the Optical Monitor on board
the ESA satellite {\it XMM-Newton} \citep{ma01}. The NGC\,346 field was 
among the selected fields observed by {\it XMM-Newton} that were 
photometrically calibrated on the basis of ground-based observations.\\

The images of this field were obtained
in April 2001 at the MPG/ESO 2.2-m telescope with the WFI mosaic
camera\footnote{
The WFI is a joint project between the European Southern Observatory
(ESO), the Max-Planck-Institut f\"ur Astronomie (MPI-A) in  Heidelberg
(Germany) and the Osservatorio Astronomico di Capodimonte (OAC) in
Naples (Italy). For more details, see http://www.ls.eso.org/lasilla/Telescopes/2p2T/E2p2M/WFI/ } 
providing a field of view of $0.5 \times 0.5$ square degrees. Due to
the unfavorable position of the SMC at this period of the year, the
seeing was less than perfect but the data  quality was still good. The
total exposure times were 25 minutes in the $U$-band and about
10 minutes in the other bands ($BVRI$). The reductions were  performed with
IRAF and our personal software. Color transformations  were done via
observations of standard Landolt fields.

\section{Source Detection and Count Rate}

The discrete X-ray sources of the field were first found by running the
CIAO wavelet algorithm {\it wavdetect} on an unbinned image of each
chip. The source detection threshold was set to $5\times 10^{-7}$,
implying that a maximum of one false source would be detected within
each chip. A total of 68 sources were then found: 7 sources in ACIS-I0,
19 sources in ACIS-I1, 16 sources in ACIS-I2, 23 sources in ACIS-I3 and
3 sources in ACIS-S3.  As {\it wavdetect} is not 100\% reliable, a second 
source detection was also performed. Using statistical thresholds 
calculated after \citet{ge86}, this second algorithm starts by eliminating 
all potential point-like sources and iteratively calculates the 
source-free background  which may vary across the field-of-view. 
Then, using these background value(s), the algorithm detects all events 
exceeding the given threshold (i.e. 3 sigma in our case), groups 
them into spatially independent sources (whose separations exceed 
2$\times$FWHM$_{PSF}$), and finally, it calculates the centroids 
of the sources. We searched for significant events on separated soft, 
medium and hard energy-band images and this second algorithm
found 13 additional sources. Five of these sources have less than 
10 net photons, and may be spurious detections. We thus decided to 
discard them. We present in 
Table~\ref{crate} the coordinates of the 75 remaining sources, listed 
by increasing Right Ascension (RA). Fig.~\ref{totalfield} shows the 
sources' positions on the ACIS-I chips. \\

Note that \hd\ is designated Src~41 and that Src~44,
belonging most probably to the extended emission, probably does not
constitute in itself a distinct point source. One source, Src~31, 
was `detected' in the emission associated with the NGC\,346 cluster.
This emission was analysed in paper I and we won't repeat the discussion 
here. Some other sources are situated near a CCD gap and their 
derived properties should therefore be treated with caution. \\

Table~\ref{crate} also lists the count rates of each source in the
$0.3-10.0$~keV energy band. The count rates were derived using the tool 
{\it dmextract}, from a spatial extraction of counts within a disk centered 
on the detected source and of radius varying from 2.5\arcsec\ to 
15\arcsec, depending of the off-axis angle. The background was extracted 
in an annulus directly surrounding the source, with the outer radius 
varying between 5\arcsec\ and 22\arcsec. The analysis of the ACIS-S 
sources was more difficult due to the high background noise of these CCDs
and the poor resolution this far off-axis. {\it wavdetect} was particularly 
confused in this region. For example, it found two close sources at the 
position of Src 75, whereas there is actually only one source. The second
algorithm did a better job, but since the derived properties of any source
in this region will not be very reliable, we decide not to analyse further
these ACIS-S sources. Note however that Src~73 seems
to possess a close neighbor (although this one was not detected by the
detection algorithms): a comparison with the PSF expected at this
position on the ACIS-S detector confirms that it may not be a single source.\\

\section{Hardness Ratios}

\begin{figure}
\plotone{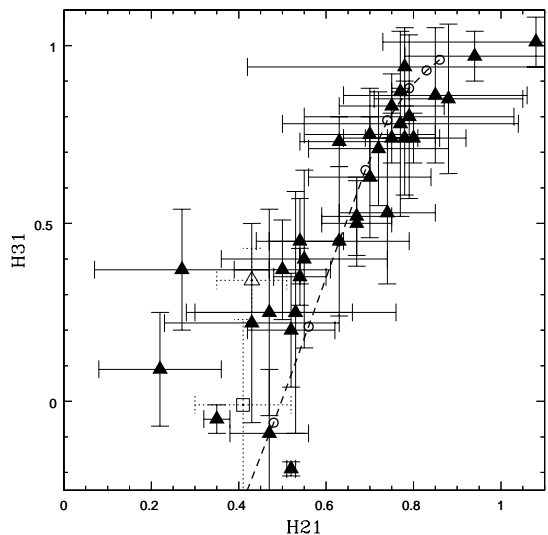}
\caption{ Hardness ratios of the brightest sources (those with a total
of $>50$~counts). The NGC\,346 cluster is shown as an open square and
\hd\ as an open triangle. Errors bars correspond to $\pm$ one standard 
deviation. The open circles connected by a dashed line represents 
the colors predicted using PIMMS for an absorbed power-law model 
with a column density
of  $4\times 10^{21}$~cm$^{-2}$ (see \S~6) and an exponent $\Gamma$
varying from 0 (upper right) to 2.8 in increments of 0.4 \citep{ir}. The
extended emission surrounding \hd\ is not shown on the diagram, being
very soft, with $H21 = -0.24\pm 0.02$ and $H31=-0.96\pm 0.05$.\label{hr}}
\end{figure}

The properties of the point sources have also been studied using their
hardness ratios (HRs). As in \citet{sa01} and \citet{bl01}, we 
have defined $S$, $M$ and $H$ as
the count rates in the $0.3-1.0$~keV, $1.0-2.0$~keV and $2.0-10.0$~keV
energy bands. We then used these data to compute two HRs: $H21$ is
defined as $(M-S)/(M+S)$ and $H31$ as $(H-S)/(H+S)$. For the
sources with more than 50~cts (i.e., a minimum count rate of $5\times
10^{-4}$~cts~s$^{-1}$), the $S$, $M$ and $H$ count rates are listed
with the corresponding HRs in Table~\ref{hrtab}. Fig.~\ref{hr} shows
$H31$ as a function of $H21$ for these sources. No supersoft source is
present : the HRs extend
mainly from $(0,0)$ to $(1,1)$, with a tight correlation between the two
HRs. The rather high HRs of the sources in the NGC\,346 field, as
compared e.g. to NGC\,4697 \citep{sa01} and NGC\,1553 \citep{bl01}, can be
explained by the larger foreground absorbing column.\\

\section{Spectral Fitting}

We have extracted the spectra of the detected point sources using the
CIAO tool $psextract$. Only sources with at
least 150~cts were fully analysed. The spectra were binned to reach 
a minimum of 10 cts per bin. Models were fitted within XSPEC, and
we used either a simple absorbed {\it mekal} model or an absorbed
power-law model, whose properties are summarized in Table~\ref{fit}.
Unfortunately, the low Signal-to-Noise of most spectra did not allow us
to discriminate between the models, so we have listed the parameters of
both models, except if they give unphysical results (e.g. $kT >> 10$~keV)
or a statistically poor fit to the data. If neither model fits the data,
other models were tested, and we have listed the parameters of the best
one. Unless otherwise stated, the abundances of the {\it mekal} models
were always fixed to the SMC mean value, $0.1 Z_{\odot}$.\\

\subsection{Absorption Column}

To better understand the spectral properties of the point sources, we
have tried to estimate the absorption column, $N(\rm H)$. We can use
two different ways. First, \hI\ surveys provide us direct
estimations of the hydrogen column: \citet{sch91} gives $4.3\times
10^{20}$~cm$^{-2}$ for the Galactic $N(\hI)$ in the direction of the SMC;
\citet{mcg82} have
measured neutral hydrogen columns in the range $(3.7-5.3)\times
10^{21}$~cm$^{-2}$ towards NGC\,346 and \citet{isr97} quotes $4.4\times
10^{21}$~cm$^{-2}$ for $N(H_2)$ towards NGC\,346. The value of the
absorbing column should then be $\sim N(\hI)_{gal}$ for sources located
in front of the SMC, 
$N(\hI)_{gal}+N(\hI)_{NGC\,346}+2N(H_2)_{NGC\,346}$ for extragalactic
sources, and intermediate values for sources within the SMC.\\

A second estimation of the extinction towards NGC\,346 can be made using
the reddening, $E(B-V)$. \citet{ma89} measured an average $E(B-V)_{SMC+gal}$ of
0.14 mag in the cluster, while \citet{sch91} estimate a Galactic
contribution of 0.07 to 0.09 mag towards the SMC. We can convert these 
values to absorption columns using the gas-to-dust ratios. For the Galaxy, 
this ratio is estimated to be $[N(\rm H)/E(B-V)]_{gal}=5.8\times 10^{21}$
cm$^{-2}$ mag$^{-1}$ \citep{bo78}. For the SMC, the quoted values are 
in the range $[N(\rm H)/E(B-V)]_{SMC}=(3.7-8.7)\times 10^{22}$~cm$^{-2}$
mag$^{-1}$ \citep{bo85,le94,fi85}. A range of typical absorption columns 
towards NGC\,346 can then be calculated using the expression 
$\left(\left[ \frac{N(\rm H)}{E(B-V)} \right]_{gal} E(B-V)_{gal}+\left[\frac{N(\rm H)}{E(B-V)}\right]_{SMC} E(B-V)_{346}\right) $. \\

\begin{figure}
\plotone{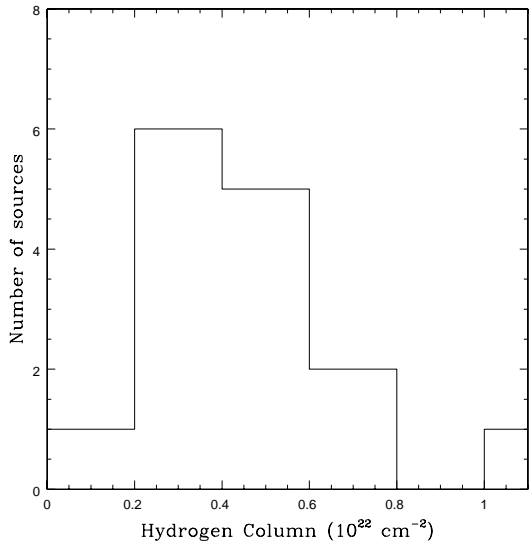}
\caption{Histogram of the different values of the absorbing column $N(\rm H)$ 
found for the point sources from the spectral fitting. Sources
with a column of $\sim 2-6\times 10^{21}$~cm$^{-2}$ are probably
associated with NGC\,346, while those sources with $N(\rm H)\geq
10^{22}$~cm$^{-2}$ are likely to be extragalactic in nature.
\label{colh}}
\end{figure}

We can thus conclude that an absorption column in the $4-20\times
10^{20}$~cm$^{-2}$ range indicates most probably a source in front of
NGC\,346 but still belonging to the SMC, while a column in the
$2-6\times 10^{21}$~cm$^{-2}$ is characteristic of a source situated in the
NGC\,346 cluster. A column higher than $\sim 10^{22}$~cm$^{-2}$
rather suggests an extragalactic source. A histogram of the values of
the absorbing columns found in our spectral fits is presented in
Fig.~\ref{colh}. It shows that most sources are indeed situated in
NGC\,346, with only few foreground sources and 
 extragalactic sources.\\

\section{Point Source Luminosity Function}

We have used the X-ray luminosities of all the point sources detected in
the field to construct a luminosity function, which can be compared to
the overall luminosity function of the SMC, as well as to other
galaxies. We thus first need to estimate the luminosities of the 
fainter sources on the basis of their count rate only. \\

To determine the countrate-to-luminosity conversion, we have
assumed that all sources lie in the SMC, at a distance of 59~kpc. 
We further assumed that the data could be well represented by
a power law of $\Gamma$=1.6. For the absorbing column, we chose 
$N(\rm H)=4\times 10^{21}$~cm$^{-2}$, an average of that 
estimated for sources belonging to the cluster in \S~5.1. 
Note that this arbitrary model appears consistent with a combined 
spectrum of all sources. For these parameters,
$PIMMS$\footnote{available on http://cxc.harvard.edu/toolkit/pimms.jsp}
gave us a typical absorbed flux of $1.4\times 10^{-11}$~erg~cm$^{-2}$~s$^{-1}$ 
for 1 cts~s$^{-1}$ in the $0.3-10.0$~keV energy band. This leads to a 
countrate-to-absorbed-luminosity conversion factor of
$5.7\times 10^{36}$~erg~cts$^{-1}$ (or $7.7\times 10^{36}$~erg~cts$^{-1}$ 
to get unabsorbed luminosities). The derived absorbed luminosities for 
the sources with $<$150 cts are listed in Table~\ref{crate}. \\

\begin{figure}
\plotone{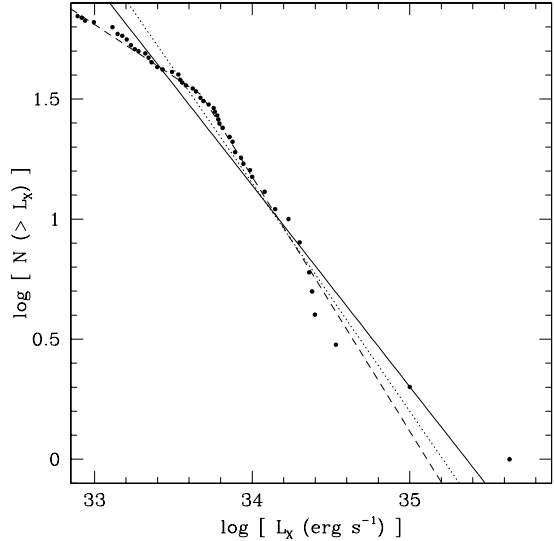}
\caption{Histogram of the cumulative X-ray luminosity function of 
all the detected sources versus $0.3-10.0$~keV unabsorbed luminosity. 
The continuous line is the best-fit power law to the data ($N \propto L_X^{-0.84}$), 
while the dotted line is the best-fit power law to the brightest data 
($N \propto L_X^{-0.95}$ for $log(L_X^{unabs})>33.5$). The dashed 
line presents the best-fit broken power law to the data, with slopes 
of $-0.43$ for $log(L_X^{unabs})<33.7$ and $-1.06$ for 
$log(L_X^{unabs})>33.7$.
\label{lognlogs}}
\end{figure}

Using this conversion factor, we also derived the unabsorbed luminosities 
of all sources, including these with known spectral fits, and we 
constructed the cumulative distribution of the sources as a function 
of the X-ray luminosity (see Fig.~\ref{lognlogs}). The best fit power 
law to our data, using a maximum likelihood minimisation 
algorithm (bayes in Sherpa v 2.3), has an exponent of $-0.84$ (with a 
90\% confidence range of $-0.77$ to $-1.32$). If we restrict the fit 
to the brightest luminosities ($log[L_X^{unabs}]>33.5]$) - since they 
are less likely to be affected by incompleteness - , the slope is 
slightly steeper, $-0.95$. However, a broken power-law provides a 
much better fit to the data. Its best fit slopes are $-0.431$ for 
$log[L_X^{unabs}]<33.68]$ and $-1.06$ for $log[L_X^{unabs}]>33.68]$ 
(with 90\% confidence ranges of $-0.42$ to $-0.44$, $-0.95$ to $-1.18$, 
and $33.60$ to $33.77$ for the two slopes and the break point, respectively).
Note that a fit using the unabsorbed luminosities from the spectral analysis 
for the brightest sources (see Table~\ref{fit}) gave very similar results.\\

For comparison, \citet{ba01} indicate that the typical trend of the 
luminosity function for the SMC, M82 and NGC\,3256 (all of which are
star-forming galaxies of varying size) is $N \propto L_X^{-0.65}$ 
for $L_X> 10^{35}$~erg~s$^{-1}$. The fit suggests that our data present 
a steeper slope than is seen for the entire SMC at higher 
luminosities, and also for other star-forming galaxies. 
But note that our observations are sampling a lower luminosity 
range than in the other quoted examples. In addition, our luminosity 
function seems actually rather complex. It presents a rather flat 
profile at low luminosities. The function then steepens for 
$L_X^{unabs}>33.7$ and flattens again for the highest 
luminosities ($L_X^{unabs}>34.5$).

\section{Source Identification}

\subsection{Comparison with Previous X-ray Observations}

We have compared our source list to the {\it ROSAT}
catalogs of SMC X-ray sources. Only 15 of the 75 sources were previously 
detected. We give in Table~\ref{ident} the correspondence between our
 sources and those previously detected  \citep{hfp}. To
be complete, we also quoted in this table the separation between our
source and the {\it ROSAT} detections, and the 90\%
confidence error on the position given in the catalog of \citet{hfp}.\\

Of these 15 sources, Src~10, 20, 69, 70, 71, 72 and 74 fall slightly 
beyond the allowed range of RA and DEC of their {\it ROSAT}
counterpart\footnote{This is also true for the {\it ROSAT} sources
observed on ACIS-S, but as all parameters derived for these sources 
are highly uncertain, we will not argue this point further.}. Either the
quoted errors were underestimated, or there are seven pairs of close
transient sources in the field, with one member in a high state and the
other in a quiescent state when the {\it ROSAT} observation was taken,
and then exactly the reverse situation when {\it Chandra} took our
data. As this possibility is very unlikely, we prefer to accept the
identification with the quoted {\it ROSAT} sources. \\

During this comparison, we also found that one previously detected
source was clearly missing: [HFP2000] 186 \citep{hfp}. [HFP2000] 186
(quoted in {\it Simbad} with an O7III counterpart even if it is very
close to a B-type star, AzV 219\footnote{AzV stars come from the catalog
by \citet{azv}}), is not far from our Src~36, but still at a separation of
56\arcsec, to be compared with the quoted positional error on this {\it
ROSAT} source, 22\arcsec.  This source may be a transient X-ray binary
(XRB). Three other {\it ROSAT} sources, [HFP2000] 202, [SHP2000] 61, 
and [SHP2000] 75 \citep{hfp,sa00}, should be present in the ACIS-S 
field, but are not clearly detected.\\

\subsection{Optical Counterparts}

To search for optical counterparts to the ACIS-I sources, we have used 
the {\it DSS}\footnote{http://archive.stsci.edu/dss/index.html} and {\it
MACHO}\footnote{http://wwwmacho.mcmaster.ca/} databases. In addition, 
we have also compared the X-ray data to
ground-based images taken with the Wide Field Imager (WFI). The limiting
magnitudes are $R\sim 17.5$~mag for the {\it DSS}, 20-21~mag for MACHO and
$\sim$21~mag for WFI. Before the comparison with the WFI images, a
global shift of 0.3 s in RA and 1.2\arcsec\ in DEC was first applied to
the coordinates of our sources, so that the position of \hd\ coincides
almost perfectly in both datasets. \\

Next we have defined an optical counterpart as a star lying less than
3\arcsec\ away from the position of the X-ray source. The results of
this search are listed in the last column of Table~\ref{crate}. 
32 sources possess at least one counterpart within the chosen
3\arcsec\ error circle, but some of these counterparts are faint 
and/or lie just at the 3\arcsec\ border. On the basis of the crowdeness
of the field, we estimate that $\sim$20\% of these identifications may be spurious detections. 15 other sources do not seem to present any 
counterpart in any of the datasets. The status of the rest of the 
sources is unclear: they present a faint counterpart in some of 
the optical images, but not in all of them, so we considered these 
identifications more dubious than
the previous ones. In Table~\ref{wfi}, we present the $UBVRI$ photometry
of the counterparts found in the WFI data, and their separation from the
X-ray source (after that the first global shift has been applied, see
above). Note that the photometry of stars with $V>$ 20~mag is rather 
uncertain, and has only an indicative value. Fig.~\ref{colmag} shows the 
color-magnitude diagram of all the sources detected in NGC\,346 by the WFI. 
The majority of the X-ray emitting sources with $V$ $<$ 20~mag are 
within the main sequence of NGC\,346.  A second group contains rather 
faint ($V>$17~mag) and redder stars which probably belongs to the SMC 
but are not physically associated with the cluster. \\

Only two counterparts in the field already have a spectral classification:
\hd\ and the star situated at the peak of the cluster's emission. 
An estimation of the spectral types of the other counterparts has been
made using the WFI photometry, the calibration of \citet{sk82}, a distance 
of 59~kpc and a 
mean reddening $E(B-V)$ of 0.14~mag \citep{ma89}. This evaluation is given 
in the last column of Table~\ref{wfi}. The majority of these newly 
cataloged counterparts seem to present a B spectral type. This result 
is not totally surprising since (a) X-ray binaries containing a Be star 
are expected to be moderately strong X-ray sources and to be brighter than 
in the case of a `normal' B-star companion, and (b) \citet{ma99} showed 
that the fraction of Be to B-type 
stars increases when metallicity decreases. With its low metallicity, the 
SMC should thus present a large population of Be/X-ray binaries, compared 
e.g. to the Galaxy, as was already suggested by e.g. \citet{ha00}. A large 
number of the X-ray sources in the field are then probably Be/X-ray binaries. 
However, accurate spectroscopic observations of these stars are needed to confirm the spectral types and firmly secure this conclusion.\\

Among the identified sources, the slightly extended X-ray source Src~2
corresponds in fact to a small group of stars, too close to each other
to disentangle their individual X-ray emission.  Src~4, 6, and 70 
are known Be/X-ray binaries (Be/XRBs). We confirm here the
previous identifications of Src~6 with [MA93] 1038\footnote{[MA93] stars
come from the catalog by \citet{ma93}} \citep{ha00} and Src~70 with a
``15-16th magnitude star'' \citep{ka96}. \citet{ha00} found another Be/XRB
candidate in our field: [HFP2000] 170 (our Src~4).  It lies between
the emission-line stars AzV 191 and [MA93] 1016.  The precise position
of this source on the {\it ROSAT} observations made the association with
either star unlikely, and we found that this source correlates well with
a faint star situated between these two bright emission-line stars. \\

\section{Source Variability}

\subsection{Short-Term Variability}

\begin{figure}
\plottwo{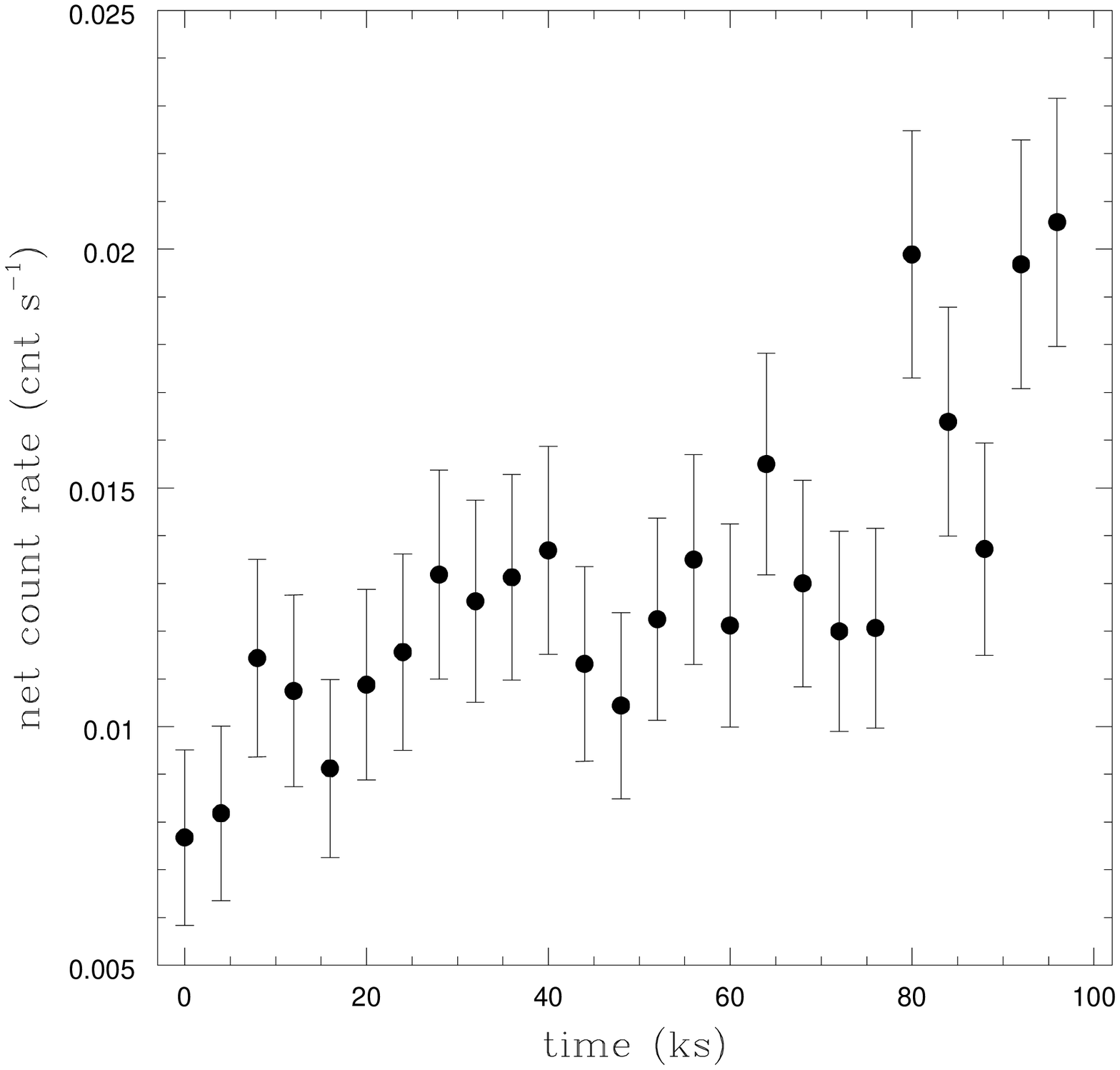}{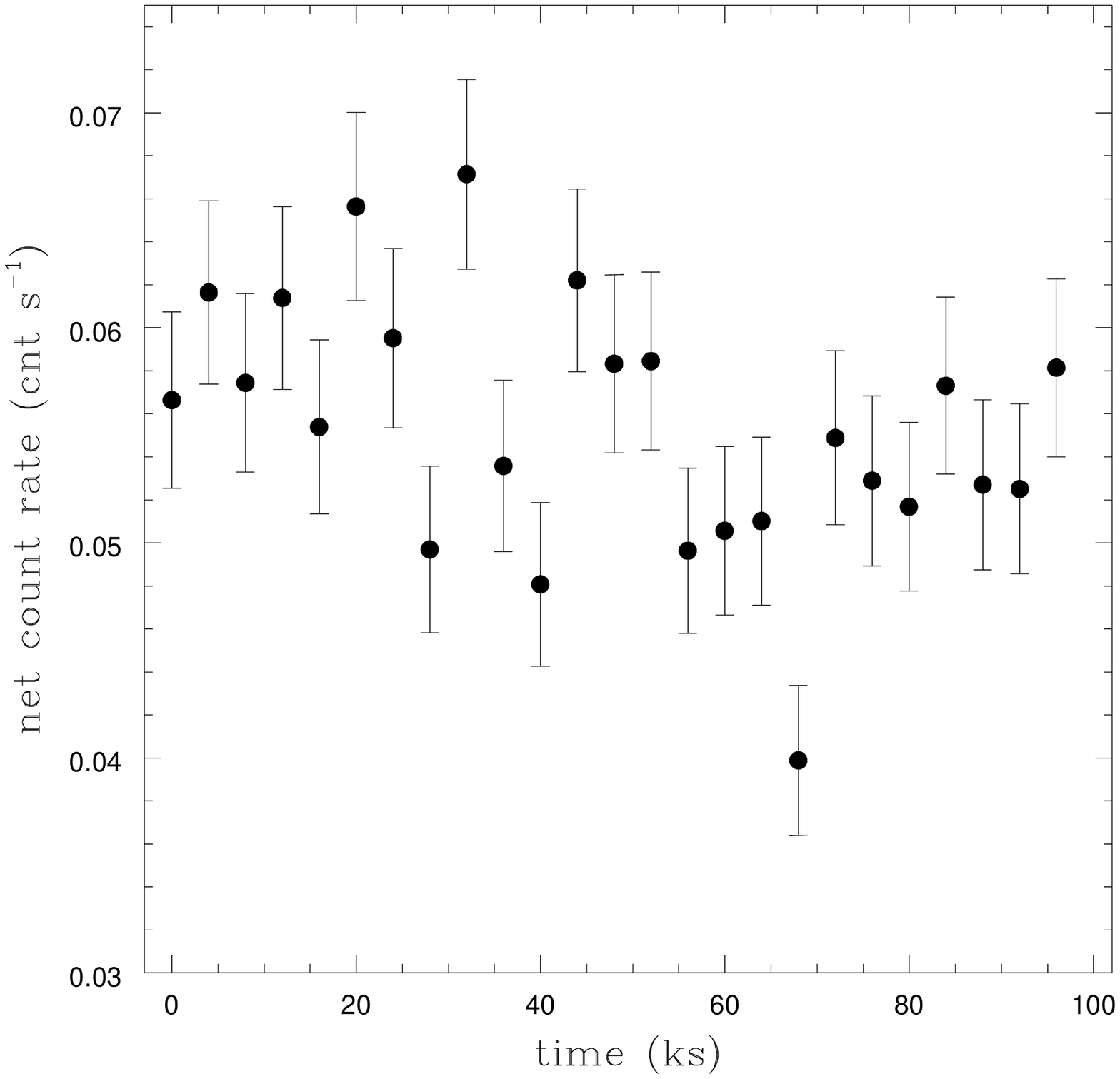}
\caption{The {\it Chandra} X-ray lightcurves of the sources identified
as being variable : Src~4 (a) and Src~6 (b). The lightcurves are in the 
0.3-10~keV range and with 25 bins of 4~ks each. A third variable source, 
Src~41 (=\hd), has been discussed extensively in paper I. The error bars
correspond to $\pm$ one standard deviation.
\label{lightc}}
\end{figure}

We have searched for short-term X-ray variability in the lightcurves of 
the brightest point sources (with $>$ 50 cts). Using a Kolmogorov-Smirnov 
test, we found that all lightcurves are consistent with a constant flux 
at the 95~\% confidence level. However, KS tests are not very sensitive, 
so we have also checked the constancy using a $\chi^2$ test. We 
have also compared the $\chi^2$ of a constant fit with the $\chi^2$ of
a linear fit, to search for any improvement when using a linear fit.
With these tests, we have detected a significant variability for Src~4, 
6, and 41 (=\hd, see paper I). Figs.~\ref{lightc}a 
and b show the lightcurves of Src~4 and 6, respectively. Src~4 and 41 
present an increase of the count rate towards the end of the exposure, 
but the variability of Src~6 seems more complex and may be related to
the Be phenomenon (see \S~7.2).\\

Considering even higher frequency variability, \citet{la02} have 
detected a period of $\sim$5~s in Src~69, the brightest point 
source of the field, which presents $\sim$6000 cts in 100~ks. 
On the basis of its luminosity and its soft spectrum, they have 
proposed it to be an Anomalous X-ray Pulsar (AXP). Using our 
photometry, we have classified the counterpart as a probable early 
B star (see Table \ref{wfi}). This suggests that the source is
in fact a Be/X-ray 
binary. From the P$_{orb}$ vs P$_{pulsar}$ diagram \citep{co86}, 
we then derived an orbital period of $\sim$25 d for this object.

\subsection{Long-Term Variability}

Long-term variations in the X-ray properties can also be checked by
comparing our {\it Chandra} data to the {\it ROSAT} PSPC observations 
of the SMC. We have already mentioned (\S~7.1) at least one {\it ROSAT} source 
which was not detected in our {\it Chandra} observation. Such variations 
are indicative of transient sources.\\

\begin{figure}
\plotone{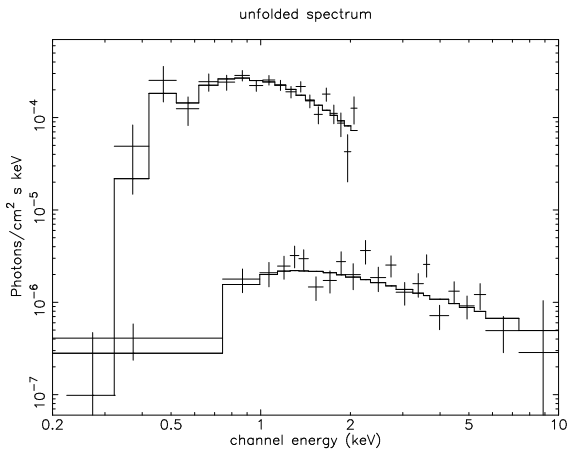}
\caption{Variations of the (unfolded) spectrum of Src~70: {\it ROSAT} data 
(upper part) extend on a smaller energy range than {\it Chandra}'s (bottom 
part) but show a substantially increased luminosity. Horizontal errors bars
correspond to the energy range covered by the bin and the vertical bars 
represent $\pm$ one standard deviation. The presented models are 
absorbed power-laws with $N(\rm H)_{ROSAT}= 0.39\times 10^{22}$~cm$^{-2}$,
$\Gamma_{ROSAT}$=2.82, $N(\rm H)_{Chandra}= 0.36\times 10^{22}$~cm$^{-2}$,
$\Gamma_{Chandra}$=1.01. \label{src67}}
\end{figure}

\begin{figure}
\plotone{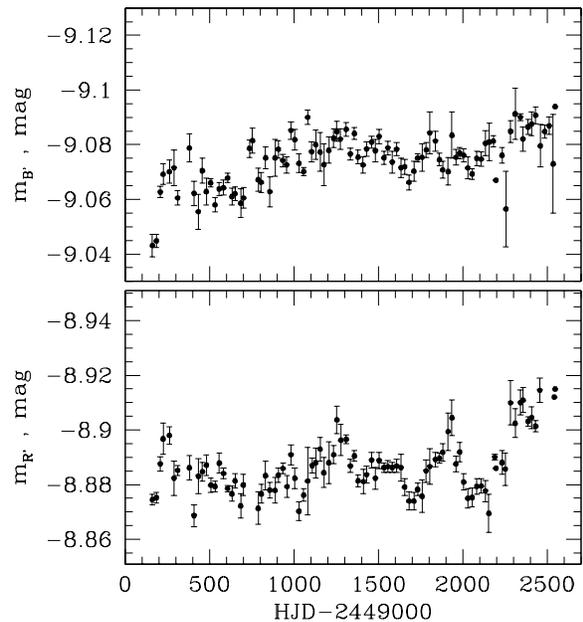}
\caption{MACHO light curve of the optical
counterpart to Src~70 in the blue and red filters, binned to 25~d. 
The last MACHO data were taken $\sim$500 d before the {\it Chandra}
observation. Error bars correspond to the dispersion of the measured
magnitudes in each bin.
\label{lc67}}
\end{figure}

One other transient source was already known in this field: Src~70.
\citet{ka96} reveal that this source is an X-ray binary evolving from
$1.3\times 10^{36}$~erg~s$^{-1}$ in the high state to $<4.6\times
10^{34}$~erg~s$^{-1}$ (i.e. undetected by $ROSAT$) six months later. 
In our dataset, this source is
in the quiescent state, since its luminosity reaches only $2\times
10^{34}$~erg~s$^{-1}$. We have used the {\it ROSAT} data to compare the
spectral properties of this source during the high and low states. The
best fit parameters to the {\it ROSAT} observation of Src~70 in the high
state are: $N(\rm H)=0.39_{0.17}^{0.68}\times 10^{22}$~cm$^{-2}$ and
$\Gamma=2.82_{1.68}^{4.22}$. The flux in the $0.3-2.0$~keV band was
$5.03\times 10^{-13}$~erg~cm$^{-2}$~s$^{-1}$ in October 1991, two orders
of magnitude larger than the flux in the same band in May 2001 ($\sim
6\times 10^{-15}$~erg~cm$^{-2}$~s$^{-1}$). The variations of the
spectrum of Src~70 can be seen in Fig.~\ref{src67}. Such luminosity
variations were also observed in other sources, e.g. 2E$0050.1-7247$,
but in this last case, the power-law steepened when the source
luminosity decreased \citep{is97}. However, if we consider only the {\it
Chandra} data in the {\it ROSAT} energy range ($0.3-2.0$~keV), the
fitted slopes of the power-laws are very similar in both datasets; thus
we cannot conclude whether the power-law has steepened or not. 
This X-ray source also possesses a variable counterpart in the
visible wavelengths.  Fig.~\ref{lc67} shows the MACHO light curve of
this counterpart in the blue and red filters, binned to 25 days. Both curves
show a long-term increase, superposed on a sinusoidal variation with a
period of $\sim$1300 d.  The properties of this interesting varying object
are certainly worth further investigation.\\

We have compared {\it Chandra} and {\it ROSAT} data in the
$0.3-2.0$~keV range for the other bright sources detected by {\it
ROSAT}. The spectral properties from the {\it ROSAT} data of these
sources are compatible with the {\it Chandra} data within the error bars.\\

Finally, we note that some of the X-ray sources may possess a
variable optical counterpart. From the MACHO database, we downloaded
the red and blue lightcurves of the possible counterparts to our X-ray 
sources. Using a CLEANed algorithm \citep{ro}, we discovered that some 
of these sources correspond to long-term variables. They are
presented in Table~\ref{macho}, along with the periods found (if
any) and their MACHO id. Except for Src~70, no variation was clearly 
detected for these sources in the X-ray domain, but future X-ray 
observations could provide additional checks.\\

\section{Summary}

In this series of articles, we report the analysis of the {\it 
Chandra} data of N66, the largest star formation region of the 
SMC. In this second article, we have focused
on the other sources detected in the field. The X-ray properties
of 75 point sources, of which 32 may possess an optical counterpart,
have been analysed. Their 
cumulative luminosity function is steeper than the global one 
of the SMC at higher luminosities. Using new photometry of the 
NGC\,346 field, we estimate that a large number of the counterparts 
are B-type stars, suggesting that many of the X-ray sources 
may be X-ray binaries. Considering their absorption column and the 
photometry of their counterparts, we also conclude that most of these 
X-ray sources probably belong to NGC\,346. Finally, due to their 
variability, some of the objects should be monitored in the future, 
both in the visible and X-ray wavebands.\\

\acknowledgments

Support for this work was provided by the National
Aeronautics and Space Administration through Chandra Award Number
GO1-2013Z issued by the Chandra X-Ray Observatory Center, which is
operated by the Smithsonian Astrophysical Observatory for and on
behalf of NASA under contract NAS8-39073.
Y.N. acknowledges support from the PRODEX XMM-OM and Integral Projects, 
contracts P4/05 and P5/36 `P\^ole d'attraction Interuniversitaire 
(SSTC-Belgium) and from PPARC for an extended visit to the University 
of Birmingham. IRS and JMH also acknowledge support from PPARC. 
AFJM thanks NSERC (Canada) and FCAR (Quebec) for financial aid.
This paper utilizes public domain data from the DSS, 2MASS, USNO,
GSC, and the MACHO Project.
This research has made use of the SIMBAD database, operated at CDS, 
Strasbourg, France and NASA's Astrophysics Data System Abstract Service.

\clearpage

\begin{figure}
\caption{{\it Chandra} ACIS-I color image of the NGC\,346 cluster. 
Three energy bands were used to create this color image : red 
corresponds to 0.3-2.0~keV, green to 2.0-4.0~keV and blue to 
4.0-10.0~keV. Before combination, the images were binned by a 
factor of two and smoothed by convolution with a gaussian of 
$\sigma$=2\arcsec. The identification numbers of the
sources found in the ACIS-I data of NGC\,346 are shown (see Table~1 for
a list of positions). The NGC\,346 cluster corresponds to Src~31 and
\hd\ is Src~41. A cross has been used to show the location of faint
sources. Note that a number of point sources were located on the 
ACIS-S3 chip and are not shown in this diagram (Srcs~72 to 75 - see 
text for more details). 
\label{totalfield}}
\end{figure}


\begin{figure}
\caption{The {\it Chandra} X-ray data superposed on the {\it DSS} image
of the field. The X-ray data were first binned by a factor of 4 and 
then smoothed by convolution with a gaussian of $\sigma=4\arcsec$.
Note that except for \hd\ and the cluster, the brightest
stars generally do not correspond to X-ray sources.
\label{dss}}
\end{figure}


\begin{figure}
\caption{The Color-Magnitude Diagram of NGC\,346 using the WFI data.
The sources listed in Table~\ref{wfi} are shown in red circles. The size of 
the circles is proportional to $\log(L_X^{abs})$ (see Table \ref{crate}). 
The error bars correspond
to the dispersion of the measured data (for some stars, the errors are 
smaller than the disk's size). Counterparts with $V>$ 20~mag 
were not included since their photometry is very uncertain.
The solid line shows an isochrone of 5~Myr for $Z$=0.004 
\citep{le01} transformed using a distance of 59~kpc and 
reddened with $R_V$=3.3 and $E(B-V)$ of 0.14~mag \citep{ma89}.  
Note that \hd\ lies at the upper left of the diagram.
\label{colmag}}
\end{figure}

\clearpage 

\begin{table}
\begin{center}
\caption{Coordinates of all detected sources with their individual count 
rate and absorbed luminosities in the $0.3-10.0$~keV band (see 
\S~6). All errors represent $\pm$ one standard deviation, with the errors 
on the coordinates given in s for the 
right ascension (RA) and in \arcsec\ for the declination (DEC). 
The last two columns indicate if the source is variable (see \S~8) and 
if it posesses an optical counterpart (see \S~7.2). A counterpart is defined 
as lying within $\sim$3\arcsec\ from the central position of the X-ray source.
A `?' status was attributed when a faint counterpart was
detected at the position, but only in one of the datasets.
\label{crate}}\medskip
\begin{tabular}{l c c c c c c} 
\tableline\tableline 
Src &RA (2000) &DEC (2000) & Count Rate& $L_X^{abs}$ & Var. ?& C.part ?\\
& $hh$ $mm$ $ss$ $\pm$ $ss$ & $^\circ$ \arcmin\ \arcsec\ $\pm$ \arcsec\ & 10$^{-4}$ cts s$^{-1}$& 10$^{33}$ erg s$^{-1}$&&\\
\tableline
 1& 00 56 52.178$\pm$0.037& $-72\ 12\ 03.80\pm 0.25$ &  26.5$\pm$2.2 & 16.9\tablenotemark{a} &N&?\\ 
 2& 00 57 12.904$\pm$0.048& $-72\ 10\ 43.25\pm 0.19$ & 8.41$\pm$1.18 & 4.83 &N&Y\\
 3& 00 57 30.060$\pm$0.070& $-72\ 10\ 09.20\pm 0.32$ & 2.93$\pm$0.90 & 1.68 &&?\\
 4& 00 57 32.411$\pm$0.012& $-72\ 13\ 01.41\pm 0.06$ &     131$\pm$4 & 44.1\tablenotemark{a} &Y&Y\\
 5& 00 57 41.568$\pm$0.043& $-72\ 09\ 02.59\pm 0.23$ & 7.57$\pm$1.17 & 4.35 &N&?\\
 6& 00 57 50.041$\pm$0.004& $-72\ 07\ 54.96\pm 0.03$ &     560$\pm$8 & 538.\tablenotemark{a} &Y&Y\\
 7& 00 57 51.162$\pm$0.055& $-72\ 08\ 27.75\pm 0.39$ & 2.13$\pm$0.71 & 1.22 &&N\\
 8& 00 57 59.561$\pm$0.071& $-72\ 16\ 19.18\pm 0.23$ & 6.36$\pm$1.14 & 3.65 &N&?\\
 9& 00 58 00.323$\pm$0.041& $-72\ 08\ 22.01\pm 0.22$ & 2.77$\pm$0.70 & 1.59 &&Y\\
10& 00 58 02.309$\pm$0.031& $-72\ 12\ 05.68\pm 0.13$ &  15.7$\pm$1.5 & 10.0\tablenotemark{a} &N&?\\
11& 00 58 07.265$\pm$0.073& $-72\ 13\ 48.43\pm 0.23$ & 1.76$\pm$0.66 & 1.01 &&Y\\
12& 00 58 08.409$\pm$0.047& $-72\ 03\ 36.98\pm 0.33$ & 8.01$\pm$1.23 & 4.60 &N&N\\
13& 00 58 09.066$\pm$0.022& $-72\ 08\ 25.62\pm 0.15$ &  10.1$\pm$1.2 & 5.80 &?&N\\
14& 00 58 09.370$\pm$0.070& $-72\ 16\ 12.20\pm 0.32$ & 4.66$\pm$1.22 & 2.68 &&N\\
15& 00 58 11.883$\pm$0.052& $-72\ 07\ 18.94\pm 0.41$ & 2.34$\pm$0.69 & 1.34 &&?\\
16& 00 58 19.900$\pm$0.070& $-72\ 16\ 17.80\pm 0.32$ & 1.76$\pm$0.76 & 1.01 &&Y\\
17& 00 58 22.688$\pm$0.042& $-72\ 14\ 48.48\pm 0.31$ & 1.65$\pm$0.68 & 0.95 &&?\\
18& 00 58 27.126$\pm$0.019& $-72\ 10\ 24.68\pm 0.16$ & 3.24$\pm$0.71 & 1.86 &&N\\
19& 00 58 27.453$\pm$0.027& $-72\ 04\ 57.36\pm 0.17$ &  15.7$\pm$1.5 & 9.55\tablenotemark{a} &N&N\\
20& 00 58 30.043$\pm$0.011& $-72\ 08\ 39.98\pm 0.05$ &  22.3$\pm$1.7 & 9.60\tablenotemark{a} &N&Y\\
21& 00 58 30.370$\pm$0.070& $-72\ 00\ 44.60\pm 0.32$ & 6.80$\pm$1.25 & 3.90 &N&N\\
22& 00 58 30.687$\pm$0.039& $-72\ 15\ 28.36\pm 0.18$ & 5.44$\pm$0.98 & 3.12 &&Y\\
23& 00 58 31.478$\pm$0.024& $-72\ 09\ 50.62\pm 0.10$ & 4.49$\pm$0.79 & 2.58 &&?\\
24& 00 58 37.336$\pm$0.020& $-72\ 03\ 21.95\pm 0.11$ &  43.4$\pm$2.3 & 22.7\tablenotemark{a} &N&?\\
25& 00 58 37.753$\pm$0.071& $-72\ 14\ 35.71\pm 0.23$ & 2.48$\pm$0.78 & 1.42 &&N\\
26& 00 58 39.386$\pm$0.034& $-72\ 02\ 27.15\pm 0.27$ &  10.2$\pm$1.3 & 5.86 &N&?\\
27& 00 58 49.750$\pm$0.070& $-72\ 17\ 14.90\pm 0.32$ & 2.99$\pm$0.98 & 1.72 &&Y\\
28& 00 58 58.200$\pm$0.070& $-72\ 01\ 07.40\pm 0.32$ & 1.08$\pm$0.69 & 0.62 &&Y\\
29& 00 59 00.744$\pm$0.008& $-72\ 13\ 28.75\pm 0.04$ &  31.8$\pm$1.9 & 18.5\tablenotemark{a} &N&Y\\
30& 00 59 03.105$\pm$0.007& $-72\ 12\ 22.80\pm 0.03$ &  30.6$\pm$1.9 & 14.6\tablenotemark{a} &N&N\\
31& 00 59 04.165$\pm$0.039& $-72\ 10\ 23.25\pm 0.15$ & \multicolumn{3}{c}{in the cluster's core, see paper I} &Y\\
32& 00 59 08.135$\pm$0.020& $-$72 12 45.85$\pm$0.10 & 1.12$\pm$0.47 & 0.64 &&?\\
33& 00 59 08.915$\pm$0.084& $-$72 15 31.01$\pm$0.22 & 1.72$\pm$0.65 & 0.99 &&?\\
34& 00 59 11.273$\pm$0.041& $-$72 12 23.61$\pm$0.16 & 1.01$\pm$0.50 & 0.58 &&?\\
\tableline
\end{tabular}
\end{center}
\end{table}
\clearpage

\setcounter{table}{0} 
\begin{table}
\begin{center}
\caption{ (continued)}\medskip
\begin{tabular}{l c c c c c c} 
\tableline\tableline
35& 00 59 12.490$\pm$0.070& $-$72 20 11.70$\pm$0.32 & 4.36$\pm$1.06 & 2.50 &&Y\\
36& 00 59 13.042$\pm$0.045& $-$72 16 17.80$\pm$0.12 &  11.4$\pm$1.3 & 6.54 &N&Y\\
37& 00 59 18.146$\pm$0.024& $-$72 11 14.90$\pm$0.10 & 2.84$\pm$0.70 & 1.63 &&?\\ 
38& 00 59 18.826$\pm$0.041& $-$72 04 23.55$\pm$0.26 & 7.33$\pm$1.10 & 4.21 &N&Y\\
39& 00 59 21.780$\pm$0.050& $-$72 15 40.12$\pm$0.19 & 2.37$\pm$0.68 & 1.36 &&Y\\
40& 00 59 25.365$\pm$0.039& $-$72 14 31.47$\pm$0.12 & 7.93$\pm$1.04 & 4.55 &N&?\\
41& 00 59 26.247$\pm$0.010& $-$72 09 52.66$\pm$0.04 &  29.8$\pm$1.9 & 13.3\tablenotemark{a} &Y&Y\\
42& 00 59 29.589$\pm$0.034& $-$72 12 34.65$\pm$0.17 & 1.12$\pm$0.47 & 0.64 &&?\\
43& 00 59 31.434$\pm$0.024& $-$72 14 17.24$\pm$0.11 &  13.4$\pm$1.3 & 7.69 &N&?\\
44& 00 59 33.762$\pm$0.042& $-$72 10 39.72$\pm$0.19 & 3.20$\pm$1.24 & 1.84 &&N\\
45& 00 59 33.972$\pm$0.018& $-$72 07 48.36$\pm$0.09 & 9.77$\pm$1.16 & 5.61 &N&Y\\
46& 00 59 34.776$\pm$0.038& $-$72 02 10.73$\pm$0.20 &  17.9$\pm$1.7 & 6.95\tablenotemark{a} &N&Y\\
47& 00 59 35.223$\pm$0.022& $-$72 08 35.08$\pm$0.10 & 3.99$\pm$0.79 & 2.29 &&?\\
48& 00 59 36.925$\pm$0.068& $-$72 16 52.46$\pm$0.17 & 5.67$\pm$1.06 & 3.25 &N&Y\\
49& 00 59 40.232$\pm$0.075& $-$72 19 03.24$\pm$0.15 &  11.0$\pm$1.4 & 6.31 &N&?\\
50& 00 59 42.910$\pm$0.045& $-$72 13 06.94$\pm$0.15 & 2.06$\pm$0.73 & 1.18 &&Y\\
51& 00 59 46.214$\pm$0.048& $-$72 02 51.73$\pm$0.33 & 6.13$\pm$1.10 & 3.52 &N&N\\
52& 00 59 47.340$\pm$0.027& $-$72 09 29.51$\pm$0.12 & 2.16$\pm$0.62 & 1.24 &&?\\
53& 00 59 49.632$\pm$0.044& $-$72 10 20.14$\pm$0.21 & 1.93$\pm$0.62 & 1.11 &&Y\\
54& 00 59 52.006$\pm$0.064& $-$72 16 40.39$\pm$0.33 & 1.69$\pm$0.69 & 0.97 &&N\\
55& 00 59 52.542$\pm$0.051& $-$72 15 30.39$\pm$0.14 & 9.32$\pm$1.14 & 5.35 &N&?\\
56& 00 59 53.254$\pm$0.054& $-$72 06 34.00$\pm$0.21 & 4.49$\pm$0.87 & 2.58 &&Y\\
57& 00 59 54.300$\pm$0.048& $-$72 10 31.79$\pm$0.18 & 1.82$\pm$0.62 & 1.04 &&Y\\
58& 01 00 06.170$\pm$0.070& $-$72 13 44.00$\pm$0.32 & 1.28$\pm$0.64 & 0.73 &&N\\
59& 01 00 11.742$\pm$0.087& $-$72 16 29.06$\pm$0.21 & 2.84$\pm$0.83 & 1.63 &&Y\\
60& 01 00 15.509$\pm$0.036& $-$72 04 41.21$\pm$0.17 &  26.4$\pm$2.0 & 12.1\tablenotemark{a} &?&Y\\
61& 01 00 17.138$\pm$0.053& $-$72 10 50.07$\pm$0.20 & 4.91$\pm$0.89 & 2.81 &&Y\\
62& 01 00 22.994$\pm$0.033& $-$72 11 28.73$\pm$0.15 &  12.6$\pm$1.4 & 7.23 &?&N\\
63& 01 00 28.081$\pm$0.055& $-$72 05 33.30$\pm$0.22 & 7.73$\pm$1.43 & 4.44 &N&?\\
64& 01 00 28.460$\pm$0.070& $-$72 04 51.50$\pm$0.32 & 2.25$\pm$0.83 & 1.29 &&?\\
65& 01 00 29.039$\pm$0.055& $-$72 05 12.33$\pm$0.26 & 6.03$\pm$1.33 & 3.46 &N&Y\\
66& 01 00 30.126$\pm$0.071& $-$72 11 29.14$\pm$0.18 & 2.23$\pm$0.66 & 1.28 &&N\\
67& 01 00 36.740$\pm$0.070& $-$72 09 13.60$\pm$0.32 & 3.45$\pm$0.87 & 1.98 &&?\\
68& 01 00 36.919$\pm$0.058& $-$72 13 16.33$\pm$0.20 & 9.32$\pm$1.29 & 5.35 &?&Y\\
69& 01 00 42.828$\pm$0.007& $-$72 11 32.36$\pm$0.02 &     606$\pm$8 & 150.\tablenotemark{a} &N&Y\\
70& 01 01 02.620$\pm$0.041& $-$72 06 57.54$\pm$0.19 &  29.2$\pm$2.0 & 25.5\tablenotemark{a} &Y&Y\\
71& 01 01 03.898$\pm$0.046& $-$72 10 04.96$\pm$0.16 &  18.0$\pm$1.6 & 5.25\tablenotemark{a} &N&Y\\
\tableline
72& 00 56 42.10$\pm$0.24& $-$72 20 27.4$\pm$1.1& \multicolumn{4}{c}{see \S 3} \\
73& 00 57 18.70$\pm$0.12& $-$72 25 30.3$\pm$1.5& \multicolumn{4}{c}{see \S 3}\\
74& 00 57 24.61$\pm$0.12& $-$72 23 58.5$\pm$1.0& \multicolumn{4}{c}{see \S 3} \\
75& 00 57 35.23$\pm$0.28& $-$72 19 30.7$\pm$2.7& \multicolumn{4}{c}{see \S 3}\\
\tableline
\end{tabular}
\tablenotetext{a}{This luminosity is an average of the luminosities from the models of Table \ref{fit}.}
\end{center}
\end{table}

\clearpage

\begin{table}
\begin{center}
\caption{Count rates in the 0.3-1.0~keV ($S$), 1.0-2.0~keV ($M$) and 
2.0-10.0~keV ($H$) energy bands, and derived hardness ratios for the 
brightest sources ($>$ 50 cts). $H21$ is defined as $(M-S)/(M+S)$ 
and $H31$ as $(H-S)/(H+S)$. Stated errors correspond to $\pm$ one 
standard deviation.
 \label{hrtab}}\medskip
\begin{tabular}{l c c c c c} 
\tableline\tableline
Src & $S$ & $M$ & $H$ & $H21$ & $H31$\\
& 10$^{-4}$ cts s$^{-1}$& 10$^{-4}$ cts s$^{-1}$& 10$^{-4}$ cts s$^{-1}$& & \\
\tableline
 1 &3.81$\pm$0.99 & 12.6$\pm$1.4 & 10.0$\pm$1.5 &0.54$\pm$0.10 &0.45$\pm$0.12\\
 2 &1.72$\pm$0.58 &2.97$\pm$0.75 &3.72$\pm$0.83 &0.27$\pm$0.20 &0.37$\pm$0.17\\
 4 & 32.7$\pm$2.0 & 68.5$\pm$2.8 & 29.5$\pm$1.9 &0.35$\pm$0.03 &$-$0.05$\pm$0.04\\
 5 &1.11$\pm$0.57 &3.87$\pm$0.78 &2.59$\pm$0.82 &0.55$\pm$0.19 &0.40$\pm$0.25\\
 6 & 23.9$\pm$1.7 &    187$\pm$4 &    350$\pm$6 &0.77$\pm$0.02 &0.87$\pm$0.01\\
 8 &1.17$\pm$0.52 &3.25$\pm$0.75 &1.94$\pm$0.84 &0.47$\pm$0.19 &0.25$\pm$0.29\\
10 &2.54$\pm$0.69 &7.55$\pm$1.01 &5.57$\pm$0.97 &0.50$\pm$0.11 &0.37$\pm$0.14\\
12 &0.27$\pm$0.42 &4.29$\pm$0.81 &3.45$\pm$0.94 &0.88$\pm$0.17 &0.85$\pm$0.21\\
13 &0.91$\pm$0.45 &6.18$\pm$0.90 &2.97$\pm$0.71 &0.74$\pm$0.11 &0.53$\pm$0.20\\
19 &0.83$\pm$0.45 &5.86$\pm$0.90 &9.04$\pm$1.15 &0.75$\pm$0.12 &0.83$\pm$0.09\\
20 &2.43$\pm$0.64 & 12.1$\pm$1.2 &7.75$\pm$1.04 &0.67$\pm$0.08 &0.52$\pm$0.11\\
21 &1.34$\pm$0.55 &3.37$\pm$0.85 &2.09$\pm$0.87 &0.43$\pm$0.20 &0.22$\pm$0.28\\
24 &6.69$\pm$1.01 & 22.7$\pm$1.7 & 14.0$\pm$1.5 &0.54$\pm$0.06 &0.35$\pm$0.08\\
26 &0.77$\pm$0.49 &4.78$\pm$0.85 &4.63$\pm$0.94 &0.72$\pm$0.16 &0.71$\pm$0.16\\
29 &2.13$\pm$0.59 & 15.2$\pm$1.4 & 14.4$\pm$1.3 &0.75$\pm$0.06 &0.74$\pm$0.07\\
30 &1.82$\pm$0.54 & 16.7$\pm$1.4 & 12.1$\pm$1.2 &0.80$\pm$0.06 &0.74$\pm$0.07\\
36 &0.84$\pm$0.47 &4.73$\pm$0.85 &5.84$\pm$0.93 &0.70$\pm$0.15 &0.75$\pm$0.13\\
38 &0.27$\pm$0.40 &3.33$\pm$0.71 &3.73$\pm$0.86 &0.85$\pm$0.21 &0.86$\pm$0.19\\
40 &0.99$\pm$0.48 &4.30$\pm$0.78 &2.64$\pm$0.65 &0.63$\pm$0.16 &0.45$\pm$0.21\\
41 &5.34$\pm$0.90 & 13.5$\pm$1.3 & 10.9$\pm$1.2 &0.43$\pm$0.08 &0.34$\pm$0.09\\
43 &0.13$\pm$0.34 &4.14$\pm$0.78 &9.10$\pm$1.13 &0.94$\pm$0.16 &0.97$\pm$0.07\\
45 &$-$0.06$\pm$0.27&1.64$\pm$0.54&8.19$\pm$1.06&1.08$\pm$0.35 &1.01$\pm$0.07\\
46 &3.14$\pm$0.76 & 10.0$\pm$1.2 &4.73$\pm$1.05 &0.52$\pm$0.10 &0.20$\pm$0.16\\
48 &0.95$\pm$0.56 &3.12$\pm$0.71 &1.60$\pm$0.71 &0.53$\pm$0.23 &0.25$\pm$0.34\\
49 &0.71$\pm$0.48 &5.61$\pm$0.92 &4.70$\pm$1.02 &0.78$\pm$0.14 &0.74$\pm$0.16\\
51 &0.33$\pm$0.42 &2.81$\pm$0.73 &2.98$\pm$0.84 &0.79$\pm$0.24 &0.80$\pm$0.23\\
55 &0.84$\pm$0.45 &4.73$\pm$0.82 &3.75$\pm$0.78 &0.70$\pm$0.14 &0.63$\pm$0.17\\
60 &2.92$\pm$0.84 & 14.8$\pm$1.4 &8.75$\pm$1.38 &0.67$\pm$0.08 &0.50$\pm$0.12\\
62 &3.36$\pm$0.79 &5.28$\pm$0.89 &4.00$\pm$0.94 &0.22$\pm$0.14 &0.09$\pm$0.16\\
63 &0.46$\pm$0.60 &3.51$\pm$0.84 &3.76$\pm$1.14 &0.77$\pm$0.27 &0.78$\pm$0.26\\
65 &$-$0.57$\pm$0.55&3.81$\pm$0.87&2.80$\pm$1.00&1.35$\pm$0.41 &1.51$\pm$0.66\\
68 &0.24$\pm$0.42 &1.89$\pm$0.64 &7.20$\pm$1.13 &0.78$\pm$0.36 &0.94$\pm$0.11\\
69 &    124$\pm$4 &    398$\pm$6 & 83.6$\pm$3.1 &0.52$\pm$0.01 &$-$0.19$\pm$0.02\\
70 &2.50$\pm$0.72 & 10.9$\pm$1.2 & 15.8$\pm$1.5 &0.63$\pm$0.09 &0.73$\pm$0.07\\
71 &3.88$\pm$0.81 & 10.9$\pm$1.2 &3.24$\pm$0.92 &0.47$\pm$0.09 &$-$0.09$\pm$0.18\\
\tableline
\end{tabular}
\end{center}
\end{table}
\clearpage

\begin{table}
\begin{center}
\caption{Parameters of the spectral fits for the sources with at least 150~cts. 
Stated errors correspond to the 90 \% confidence level.
Abundances are fixed at $0.1\,Z_{\odot}$, except for src~69, 
where the best fit is found for $Z=0.0022_{0.0001}^{0.0094} Z_{\odot}$.
Absorbed luminosities are given in the spectral range
$0.3-10$~keV and for a distance of 59~kpc. \label{fit}}\medskip 
\begin{tabular}{l l c c c c c} 
\tableline\tableline
Src & Model & $L_X^{abs}$  & $\chi^2_{\nu}$& $N(\rm H)$ & Other Parameters & $N_{dof}$\\
 &   & 10$^{34}$ ergs s$^{-1}$ & &  10$^{22}$ cm$^{-2}$ & ($kT$ in keV) & \\
\tableline
1 & wabs*mekal & 1.65 & 0.81 & 0.29$_{0.15}^{0.49}$ & $kT=9.37_{4.56}^{55.75}$ & 29  \\
 & wabs*pow &    1.72 & 0.82 & 0.36$_{0.21}^{0.57}$ & $\Gamma$=1.65$_{1.32}^{1.98}$ & 29 \\
4 & wabs*mekal & 4.26 & 1.19 & 0.17$_{0.13}^{0.19}$ & $kT=2.87_{2.50}^{3.60}$ & 97  \\
 & wabs*pow &    4.55 & 1.06 & 0.29$_{0.24}^{0.32}$ & $\Gamma$=2.36$_{2.20}^{2.51}$ & 97 \\
6 & wabs*pow   & 55.0 & 1.57 & 0.51$_{0.45}^{0.56}$ & $\Gamma$=0.86$_{0.80}^{0.93}$ & 317  \\
 & wabs*(mekal+&52.5 & 1.52 & 1.02$_{0.89}^{1.20}$ & $kT=0.58_{0.49}^{0.65}$ $N_{H2}$=0.87$_{0.17}^{1.64}$ & 314 \\
 & wabs*pow) &  & & & $\Gamma$=1.20$_{1.10}^{1.37}$ &  \\
10 & wabs*pow  & 1.00 & 0.91 & 0.16$_{0.04}^{0.34}$ & $\Gamma$=1.50$_{1.18}^{1.94}$ & 14 \\ 
19 & wabs*mekal& 0.95 & 1.86 & 1.39$_{0.74}^{2.13}$ & $kT=5.52_{2.04}^{100.00}$ & 14 \\
 & wabs*pow    & 0.96 & 1.78 & 1.80$_{0.87}^{2.89}$ & $\Gamma$=2.21$_{1.68}^{3.16}$ & 14  \\
20 & wabs*mekal& 0.92 & 0.30 & 0.50$_{0.35}^{0.65}$ & $kT=3.68_{2.57}^{6.60}$ & 19 \\ 
 & wabs*pow    & 1.00 & 0.24 & 0.64$_{0.48}^{0.91}$ & $\Gamma$=2.18$_{1.85}^{2.40}$ & 19 \\
24 & wabs*mekal& 2.22 & 1.10 & 0.19$_{0.12}^{0.27}$ & $kT=7.43_{4.52}^{16.60}$ & 39 \\
 & wabs*pow    & 2.32 & 1.06 & 0.26$_{0.19}^{0.36}$ & $\Gamma$=1.75$_{1.52}^{1.92}$ & 39  \\ 
29 & wabs*pow  & 1.85 & 0.94 & 0.54$_{0.39}^{0.73}$ & $\Gamma$=1.64$_{1.38}^{1.86}$ & 29  \\
30 & wabs*mekal& 1.42 & 0.61 & 0.57$_{0.44}^{0.74}$ & $kT=4.24_{3.63}^{6.66}$ & 25 \\
 & wabs*pow    & 1.50 & 0.61 & 0.74$_{0.57}^{0.95}$ & $\Gamma$=2.16$_{1.82}^{2.30}$ & 25  \\
41 & wabs*mekal& 1.30 &1.24 & 0.22$_{0.14}^{0.35}$ & $kT=7.04_{4.35}^{13.35}$ & 27   \\  
 & wabs*pow    & 1.36 &1.30 & 0.28$_{0.19}^{0.44}$ & $\Gamma$=1.74$_{1.53}^{1.89}$ & 27  \\
46 & wabs*mekal& 0.67 &1.26 & 0.39$_{0.26}^{0.58}$ & $kT=2.60_{1.29}^{4.47}$ & 17   \\  
 & wabs*pow    & 0.72 &1.23 & 0.56$_{0.41}^{0.82}$ &
$\Gamma$=2.51$_{2.09}^{3.10}$ & 17  \\
60 & wabs*mekal& 1.15 &0.81 & 0.40$_{0.33}^{0.58}$ & $kT=4.06_{2.46}^{8.63}$ & 27   \\  
 & wabs*pow    & 1.27 &0.78 & 0.51$_{0.35}^{0.77}$ & $\Gamma$=2.05$_{1.71}^{2.46}$ & 27  \\
69\tablenotemark{a} & wabs*mekal& 14.9 &1.42 & 0.41$_{0.38}^{0.44}$ & $kT=1.04_{0.99}^{1.11}$ & 163 \\  
 & wabs*pow    & 15.1 &1.75 & 0.71$_{0.67}^{0.75}$ & $\Gamma$=3.82$_{3.71}^{3.93}$ & 164  \\
70 & wabs*pow  & 2.55 &1.13 & 0.36$_{0.20}^{0.63}$ & $\Gamma$=1.01$_{0.76}^{1.32}$ & 29 \\
71 & wabs*mekal& 0.51 &1.45 & 0.25$_{0.15}^{0.50}$ & $kT=1.91_{0.99}^{3.12}$ & 16 \\  
 & wabs*pow    & 0.54 &1.27 & 0.47$_{0.30}^{0.82}$ & $\Gamma$=2.93$_{2.44}^{3.31}$ & 16  \\
\tableline
\end{tabular}
\tablenotetext{a}{$Z=0.0022_{0.0001}^{0.0094} Z_{\odot}$}
\end{center}
\end{table}
\clearpage

\begin{table}[htb]
\begin{center}
\caption{ X-ray sources previously detected by ROSAT in the NGC\,346 
field. The ids and positions of the [HFP2000] sources are taken 
from \citet{hfp}. The last columns list the positional error of the 
ROSAT sources (from Haberl et al. 2000), and the separation between 
{\it Chandra} and ROSAT sources. \label{ident}}\medskip
\begin{tabular}{l c c c c c} 
\tableline\tableline
 Src & [HFP2000] & RA (2000) & DEC (2000) & Pos. Er.& Sep. \\ 
&  & $hh$ $mm$ $ss$ & $^\circ$ \arcmin\ \arcsec\ & \arcsec\ & \arcsec \\
\tableline
1 &  164 & 00 56 52.4 & $-$72 12 00 &  9.8 &  3.9\\
4 &  170 & 00 57 31.8 & $-$72 13 03 &  3.6 &  3.2\\
6 &  136 & 00 57 50.1 & $-$72 07 56 &  5.1 &  1.1\\
10&  165 & 00 58 00.8 & $-$72 12 09 &  6.3 &  7.7\\
11&  173 & 00 58 01.7 & $-$72 13 51 & 28.2 & 25.6\\
16&  185 & 00 58 16.3 & $-$72 16 18 & 19.0 & 16.4\\
20&  139 & 00 58 28.1 & $-$72 08 48 & 11.0 & 12.0\\
24&  116 & 00 58 35.2 & $-$72 03 12 & 19.4 & 14.0\\
60&  123 & 01 00 13.1 & $-$72 04 40 & 15.9 & 11.2\\
69&  162 & 01 00 41.5 & $-$72 11 34 &  2.3 &  6.3\\
70&  132 & 01 01 01.1 & $-$72 06 57 &  2.8 &  7.0\\
71&  150 & 01 01 07.9 & $-$72 10 26 & 19.0 & 27.9\\
\tableline
72&  206 & 00 56 39.8 & $-$72 20 28 &  7.4 &  10.5\\
73&  234 & 00 57 18.9 & $-$72 25 30 &  3.4 &  1.0\\
74&  223 & 00 57 21.9 & $-$72 23 57 & 11.6 &  12.4\\
\tableline
\end{tabular}
\end{center}
\end{table}
\clearpage

\begin{table}
\begin{center}
\caption{Photometry of the optical counterparts (WFI data), and separation 
between the X-ray source and its counterpart. The error quoted in the 
$\sigma_{V}$ column represents the dispersion of the measured data. A `:' 
denotes an uncertain value. If the counterpart is cataloged, the identifier 
is given in the last column of the Table, together with its spectral type in 
parentheses, if known. Identifiers beginning by `S010202' are taken from the 
Guide Star Catalog-II. When it was possible, an estimation of the spectral 
type of the source was made, assuming that it belongs to NGC\,346 
($E(B-V)$=0.14~mag, d=59~kpc). The result is written in italics in the last column.
\label{wfi}}\medskip
\begin{tabular}{l r r r r r r c c l} 
\tableline\tableline
Src & $V$ & $B-V$ & $U-B$ & $V-R$ & $R-I$ & $V-I$ & $\sigma_{V}$ & Sep. & Remarks\\
&(mag)&(mag)&(mag)&(mag)&(mag)&(mag)& (mag)& (\arcsec)& \\ 
\tableline
1 & 21.32 & 0.55 &  & 0.43 & 2.14 & 2.57 & 0.14 & 1.6 & {\it F5III}\\
2\tablenotemark{a} & 17.14 & 0.54 & & 0.53 & 0.29 & 0.82 & 0.02 & 2.2 & S010202097601, \\
 & & & & & & & & & USNO-A2.0 0150-00623029\\
3 & 19.67 & $-0.01$ &  & $-0.08$ & $-0.22$ & $-0.30$ & 0.01 & 0.9 & {\it B9V}, S0102029846 \\
4 & 18.19 & 0.50 &  & 0.16 &  &  & 0.05 & 0.9 & S01020206684, \\
 & & & & & & & & & USNO-A2.0 0150-00629718\\
6 & 15.71 & 0.01 & $-0.96$ & 0.10 & 0.05 & 0.14 & 0.01 & 0.5&[MA93]1038\tablenotemark{b}\\
 & & & & & & & & & {\it late O, early B star}\\ 
9 & 18.86 & $-0.12$ & $-0.49$ & 0.09 &  &  & 0.09 & 1.1 & {\it B6V}, S010202012821,\\
 & & & & & & & & &  USNO-A2.0 0150-00638074\\
10 & 17.57 & $-0.04$ & $-0.72$ & $-0.10$ &  &  & 0.02 & 1.8 & {\it B2-3V}, S0102027728\\
11\tablenotemark{c} & 19.11 & 0.25 &  & 0.62 & 1.46 & 2.08 & 0.04 & 1.3 & S01020205791, \\
 & & & & & & & & & 2MASS 0058077-721349\\
16 & 20.70 & $-0.12$ &  & $-0.31$ &  &  & 0.05 & 0.2 & \\
20 & 20.88: & 1.28: &  & 1.15: & 0.86: & 2.01: & 0.27 & 0.7 & \\
22 & 19.25 & $-0.16$ & $-0.36$ & $-0.13$ & 1.00 & 0.87 & 0.10 & 0.7 & \\
26 & 20.05 & $-0.08$ &  & $-0.24$ &  &  & 0.04 & 0.1 & \\
28 & 18.71 & 1.49 &   & 0.95 & 1.37 & 2.33 & 0.06 & 2.5 & {\it K4III}, S010202028457 \\
29 & 19.41 & 0.16 &  &  &  & 0.53 & 0.08 & 1.7 & {\it A1III}\\
31 & 12.61 & $-0.15$ & $-1.07$ & $-0.06$ & $-0.18$ & $-0.24$ & 0.01 & 0.5 & MPG435 (O4III(f))\tablenotemark{e}\\
34\tablenotemark{c} & 19.55 & $-0.04$ &  & $-0.28$ &  &  & 0.11 & 2.0 & {\it B9V}\\
36\tablenotemark{c} & 19.29 & 0.10 & $-0.46$ &  &  &  & 0.10 & 1.7 & 2MASS 0059123-721617\\
37 & 18.73: & 0.80: &  & 0.23: & 0.62: & 0.85: & 0.15 & 0.6 & \\
39 & 20.57: & 0.73: & & $-0.76$: & 2.36: & 1.60: & 0.42 & 2.3 & \\
41 & 11.32 & $-0.32$ & $-0.88$ & 0.13 & $-0.10$ & 0.03 & 0.01 & 0.3 & HD 5980 (LBV)\\
45\tablenotemark{f} & 21.24: & 0.45: &  & 0.07: & 0.16: & 0.24: & 0.64 & 0.4 & \\
46 & 19.42 & 0.55 & 0.00: & 0.46 & 0.52 & 0.98 & 0.07 & 0.9 & \\
\tableline
\end{tabular}
\end{center}
\end{table}
\clearpage

\setcounter{table}{4}
\begin{table}[phtb]
\begin{center}
\caption{(continued)}\medskip
\begin{tabular}{l r r r r r r c c l} 
\tableline
48 & 19.54: & 0.87: &  & 0.19: & 0.72: & 0.91: & 0.21 & 1.1 & \\
50 & 19.42 & 2.19 &  & 0.88 & 1.67 & 2.55 & 0.09 & 0.9 & 2MASS 0059431-721307\\
53\tablenotemark{f} & 20.41 & $-0.11$ & $-0.65$: &  &  &  & 0.13 & 1.4 & \\
56\tablenotemark{f} & 20.86 & 0.25: &  & 0.60 & 0.38: & 0.98: & 0.05 & 1.6 & {\it A3V}\\
57 & 18.08 & 0.05 & 0.13 & 0.08 &  &  & 0.01 & 1.3 & \\
59\tablenotemark{f} & 19.26 & 0.60 & $-0.29$ & 0.46 &  &  & 0.10 & 1.1 &  \\
60\tablenotemark{f} & 21.32: & 1.07: &  & 0.83: & 1.08: & 1.91: & 0.23 & 1.1 & \\
61\tablenotemark{a} & 18.87 & $-0.07$ &  & 0.42 &  &  & 0.08 & 1.2 & {\it B7-8V}\\
63\tablenotemark{a} & 20.82: & 0.32: & $-3.60$: & $-0.22$: & 1.42: & 1.20: & 0.22 & 2.0 & \\
64 & 21.03: & & & & & & 0.39 & 1.0 & \\ 
65 & 19.55 & 0.88 &  & 0.32 &  &  & 0.09 & 1.3 & \\
68 & 16.73 & $-0.11$ & $-0.84$ & $-0.09$ & $-0.22$ & $-0.31$ & 0.01 & 2.9 & {\it B2V}, S010202096809, \\
 & & & & & & & & & 2MASS 0100370-721315\\
69\tablenotemark{g} & 17.82 & $-0.13$ & $-0.68$ & $-0.06$ & $-0.06$ & $-0.12$  & 0.02 & 1.4 & {\it B3V}, S01020208968\\
70 & 15.81 & 0.00 & $-0.95$ & 0.08 & 0.05 & 0.14 & 0.02 & 0.6 & [MA93]1240\tablenotemark{h} \\
 & & & & & & & & &{\it late O, early B star}\\
71\tablenotemark{g} & 18.42 & 0.05 &  & 0.24 & $-0.02$ & 0.23 & 0.10 & 2.3 & {\it B6V}, S010202010679\\
\tableline
\end{tabular}
\tablenotetext{a}{closest within a group}
\tablenotetext{b}{=S010202013449, 2MASS 0057504-720756, USNO-A2.0 0150-00634949}
\tablenotetext{c}{northern component of double}
\tablenotetext{d}{=S01020202934, 2MASS 0058502-721713}
\tablenotetext{e}{=2MASS 0059045-721024, cluster's core}
\tablenotetext{f}{faint}
\tablenotetext{g}{brightest in a group}
\tablenotetext{h}{=S010202016442, 2MASS 0101029-720658}
\end{center}
\end{table}
\clearpage

\begin{table}[htb]
\begin{center}
\caption{Periods and separations of the optical variable counterparts 
found in the MACHO database. In the last column, a letter in parentheses
indicates in which band the period was found: `B' refers to the blue 
lightcurve, and `R' to the red one.\label{macho}}\medskip 
\begin{tabular}{l l c l} 
\tableline\tableline
 Src & MACHO id & Sep. & Period(s)\\ 
\tableline
11\tablenotemark{a}& 207.16490.238&1.56"&357 d (R \& B)\\
  & 207.16490.370&1.68"&357 d (B)\\
27& 207.16546.22&2.57"&9.12 d (R \& B)\\
48\tablenotemark{b}& 207.16547.618&1.90"&227 d (R), no data in B \\
68& 206.16661.96&1.35"&long-term variable (B)\\
70& 206.16663.16&1.13"&$\sim$1300 d (R \& B)\\
\tableline
\end{tabular}
\tablenotetext{a}{Suspiciously close to $P$=1 yr.}
\tablenotetext{b}{Unusual lightcurve : the scatter increases towards the maximum.}
\end{center}
\end{table}
\end{document}